\documentclass[12pt]{amsart}

\newcommand{\HH}{\mathcal H}
\newcommand{\R}{\mathbb R}
\newcommand{\x}{\mathbf x}
\newcommand{\D}{\mathcal D}
\newcommand{\C}{\mathcal C}

\begin{document}

\title[Generalized Schrodinger eqn and quantum field theory]
      {Generalized Schrodinger equation and
       constructions of quantum field theory}
\author{A. V. Stoyanovsky}
\address{Moscow Center for Continuous Mathematical Education,
Bolshoj Vlasjevskij per., 11, Moscow, 119002, Russia}

\begin{abstract}
The generalized Schrodinger equation deduced in the earlier papers
is compared with conventional constructions of quantum field theory.
In particular, it yields the usual Schrodinger equation of quantum
field theory written without normal ordering. This leads to a definition
of certain mathematical version of Feynman integral.
\end{abstract}

\maketitle

In the papers [2,3] a generalization of the Schrodinger equation to
multidimensional variational problems has been deduced. In the present
paper this equation is compared with constructions of quantum field theory.
To make the paper more self-contained, first
the derivation of the generalized Schrodinger equation
from [3] is briefly recalled.

\subsection*{1. Derivation of the generalized Schrodinger equation}

Consider the action functional of the form
\begin{equation}
J=\int_D F(x^0,\ldots,x^n,z^1,\ldots,z^m,z^1_{x^0},\ldots,z^m_{x^n})\,
dx^0\ldots dx^n.
\end{equation}
Here $x^0,\ldots,x^n$ are the independent variables,
$z^1,\ldots,z^m$ are the dependent variables,
$z^i_{x^j}=\frac{\partial z^i}{\partial x^j}$, and integration goes over
an $(n+1)$-dimensional surface $D$ in $\R^{m+n+1}$ (the graph of the
functions $z^i(x)$) with the boundary $\partial D$. Suppose that for each
$n$-dimensional surface $C$ in $\R^{m+n+1}$ given by the equations
\begin{equation}
x^j=x^j(s^1,\ldots,s^n),\ \ z^i=z^i(s^1,\ldots,s^n)
\end{equation}
and sufficiently close
to a fixed $n$-dimensional surface, there exists a unique $(n+1)$-dimensional
surface $D$ with the boundary $\partial D=C$ which is an extremal of the
variational problem, i.e., the graph of a solution $z(x)$ to the
Euler--Lagrange equations. Denote by $S(C)$ the integral (1)
over the surface $D$. Then one has the following well known formula for the
variation of the functional $S$:
\begin{equation}
\delta S=\int_C \left(\sum p^i\delta z^i-\sum H^j\delta x^j\right)\,
ds^1\ldots ds^n,
\end{equation}
or
\begin{equation}
\begin{aligned}{}
\frac{\delta S}{\delta z^i(s)}&=p^i(s),\\
\frac{\delta S}{\delta x^j(s)}&=-H^j(s),
\end{aligned}
\end{equation}
where
\begin{equation}
\begin{aligned}{}
p^i&=\sum_l(-1)^lF_{z^i_{x^l}}
\frac{\partial(x^0,\ldots,\widehat{x^l},\ldots,x^n)}{\partial(s^1,\ldots,s^n)},\\
H^j&=\sum_{l\ne j}(-1)^lF_{z^i_{x^l}}z^i_{x^j}
\frac{\partial(x^0,\ldots,\widehat{x^l},\ldots,x^n)}{\partial(s^1,\ldots,s^n)}\\
&+(-1)^j(F_{z^i_{x^j}}z^i_{x^j}-F)
\frac{\partial(x^0,\ldots,\widehat{x^j},\ldots,x^n)}{\partial(s^1,\ldots,s^n)}.
\end{aligned}
\end{equation}
Here $\frac{\partial(x^1,\ldots,x^n)}{\partial(s^1,\ldots,s^n)}=\left|
\frac{\partial x^j}{\partial s^i}\right|$ is the Jacobian;
the cap over a variable means that the variable is omitted; and the summation
sign over the index $i$ repeated twice is omitted.

The quantities $p^i$ and $H^j$ satisfy the relations
\begin{equation}
p^iz^i_{s^k}-H^jx^j_{s^k}=0,\ \ \ k=1,\ldots,n,
\end{equation}
and one more relation which we denote by
\begin{equation}
\HH(x^j(s),z^i(s),x^j_{s^k},z^i_{s^k},p^i(s),-H^j(s))=0.
\end{equation}
From these relations one can express (in general) the quantities $H^j$
as functions of $p^i$:
\begin{equation}
H^j=H^j(x^l,z^i,x^l_{s^k},z^i_{s^k},p^i),\ \ \ j=0,\ldots,n.
\end{equation}
Substituting (4) into (6,7) or into (8), one obtains the generalized
Hamilton--Jacobi equations:
\begin{equation}
\begin{aligned}{}
&\frac{\delta S}{\delta z^i(s)}z^i_{s^k}+\frac{\delta S}{\delta x^j(s)}
x^j_{s^k}=0,\ \ \ k=1,\ldots,n,\\
\HH&\left(x^j,z^i,x^j_{s^k},z^i_{s^k},
\frac{\delta S}{\delta z^i(s)},\frac{\delta S}{\delta x^j(s)}\right)=0,
\end{aligned}
\end{equation}
or
\begin{equation}
\frac{\delta S}{\delta x^j(s)}+H^j\left(x^l,z^i,x^l_{s^k},z^i_{s^k},
\frac{\delta S}{\delta z^i(s)}\right)=0,\ \ \ j=0,\ldots,n.
\end{equation}
The first $n$ equations in the system (9) correspond to the fact that the
functional $S(C)$ is independent on the parameterization of the surface $C$.
For a theory of the generalized Hamilton--Jacobi equations, see [3] or [1].

Assume that functions (8) are polynomials in the variables $p^i$. Let us
make the following substitution in the generalized Hamilton--Jacobi equations:
\begin{equation}
\begin{aligned}{}
&\frac{\delta S}{\delta x^j(s)} \to -ih\frac{\delta}{\delta x^j(s)},\\
&\frac{\delta S}{\delta z^{i'}(s)}=p^{i'} \to -ih\frac{\delta}{\delta z^{i'}(s)}.
\end{aligned}
\end{equation}
Here $i$ is the imaginary unit, $h$ is a very small constant (the Plank constant).
We obtain the system of linear variational differential equations which can be
naturally called the generalized Schrodinger equations:
\begin{equation}
-ih\frac{\delta \Psi}{\delta x^j(s)}+H^j\left(x^l,z^{i'},x^l_{s^k},z^{i'}_{s^k},
-ih\frac{\delta}{\delta z^{i'}(s)}\right)\Psi=0,\ \ \ j=0,\ldots,n,
\end{equation}
or
\begin{equation}
\begin{aligned}{}
&\frac{\delta \Psi}{\delta z^{i'}(s)}z^{i'}_{s^k}+\frac{\delta \Psi}{\delta x^j(s)}
x^j_{s^k}=0,\ \ \ k=1,\ldots,n,\\
\HH&\left(x^j,z^{i'},x^j_{s^k},z^{i'}_{s^k},
-ih\frac{\delta}{\delta z^{i'}(s)},-ih\frac{\delta}{\delta x^j(s)}
\right)\Psi=0
\end{aligned}
\end{equation}
provided that the left-hand side of equation (7) is also a polynomial in $p^i$
and $H^j$. Here $\Psi=\Psi(C)$ is the unknown complex-valued functional of
the surface $C$ (2). The first $n$ equations in the system (13) mean that the
value $\Psi(C)$ is independent on the parameterization of the surface $C$.

\subsection*{2. Comparison with quantum field theory}

Since the value $\Psi(C)$ is independent of the parameterization of the surface $C$,
we can choose a particular parameterization. Put $s^1=x^1,\ldots,s^n=x^n$. The
generalized Schrodinger equation becomes a single equation, which we choose to be
the first equation from the system (12) corresponding to $j=0$. This equation
can be easily computed from (5):
\begin{equation}
-ih\frac{\delta \Psi}{\delta x^0(\x)}+H\left(x^0(\x),\x,z^i(\x),
\frac{\partial x^0}{\partial\x},\frac{\partial z^i}{\partial\x},
-ih\frac{\delta}{\delta z^i(\x)}\right)\Psi=0,
\end{equation}
where $\x=(x^1,\ldots,x^n)$, $\frac{\partial z^i}{\partial\x}=
(\frac{\partial z^i}{\partial x^1},\ldots,
\frac{\partial z^i}{\partial x^n})$, and
\begin{equation}
\begin{aligned}{}
H&=H\left(x^0,\x,z^i,\frac{\partial x^0}{\partial\x},
\frac{\partial z^i}{\partial\x},p^i\right)
=\sum_i p^iz^i_{x^0}-F(x^0,\x,z^i,z^i_{x^0},z^i_{x^j}),\\
z^i_{x^j}&=\frac{\partial z^i}{\partial x^j}-z^i_{x^0}
\frac{\partial x^0}{\partial x^j},\ \ \ j=1,\ldots,n,\\
p^i&=F_{z^i_{x^0}}-\sum_{j=1}^n F_{z^i_{x^j}}\frac{\partial x^0}{\partial x^j}
=\frac{\partial F}{\partial z^i_{x^0}}.
\end{aligned}
\end{equation}
That is, $H$ is the Legendre transform of the Lagrangian $F$ with respect to the
variables $z^i_{x^0}$. Equation (14) looks approximately like the
Tomonaga--Schwinger equation [4], with three differences: a) it has a mathematical
sense, unlike the Tomonaga--Schwinger equation [5]; b) instead of Hilbert space
of states, we have the space of functionals of functions $z^i(\x)$ for any
spacelike surface $x^0=x^0(\x)$; c) there are no normal orderings.
Given any functional $\Psi_0(z^i(\x))$ for a fixed spacelike surface $x^0=x^0(\x)$,
equation (14) describes the evolution of the functional $\Psi_0$ as the
spacelike surface varies. We conjecture that equation (14) is integrable,
i.e., the result of evolution exists (for conventional Lagrangians)
and depends only on the initial and final spacelike
surfaces and not on the concrete way of evolution of the surface.

The following definition is motivated in part by the geometric picture
of excitations propagating along surfaces from [3].

{\bf Definition.} Assume we are given: 1)~two spacelike surfaces
$\C_0$: $x^0=x^0(\x)$ and $\C_1$ bounding a domain $\D$ in $\R^{n+1}$;
2)~a functional $\Psi_0(z^i(\x))$ on the space of functions $z^i(\x)$ on the
first surface $\C_0$;
and  3)~functions $z^i=z^i_1(\x)$ on the second surface $\C_1$.
Then {\it a mathematical version of the Feynman integral}
\begin{equation}
\int e^{\frac{i\int_{\D} F\,dx}{h}}\Psi_0(z^i(x^0(\x),\x)) \prod Dz^i(x),
\end{equation}
taken over all functions $z^i(x)$ on the closed domain $\D$ whose
restriction to the second boundary surface $\C_1$ coincides with the given
functions $z^i_1$ (and restriction to $\C_0$ is arbitrary), is defined as
follows. Take the functional $\Psi_0$ as the initial value corresponding
to the first surface $\C_0$, and consider the evolution of this functional
given by the generalized Schrodinger equation (14). We obtain a functional
$\Psi$ of a spacelike surface and of functions $z^i$ on it. Take the value
of this functional at the surface $\C_1$ and the functions $z^i_1$. This is the
desired Feynman integral.

\medskip

Considering evolution of flat spacelike surfaces
$x^0(\x)=const=t$, we arrive
at the evolution equation for a functional $\Psi(t,z^i(\x))$:
\begin{equation}
ih\frac{\partial\Psi}{\partial t}=\int H\left(t,\x,z^i(\x),
\frac{\partial z^i}{\partial\x},-ih\frac{\delta}{\delta z^i(\x)}\right)
\Psi\,d^n\x.
\end{equation}
This equation differs from the usual quantum field theory Schrodinger equation
by absence of normal orderings; but it has a mathematical sense.

It is natural to ask whether the operator
of evolution of the functional $\Psi$ from $t=-\infty$ to $t=\infty$
described by equation (17) exists for conventional Lagrangians of
quantum field theory.

\end{document}